\documentclass[aps,prb,superscriptaddress,twocolumn,showpacs,10pt]{revtex4-1}
\usepackage{amsmath}
\usepackage{amssymb}
\usepackage{graphicx}
\usepackage{array}
\usepackage{dcolumn}
\usepackage{subfigure}
\usepackage{url}
\usepackage{color}
\usepackage{float}
\usepackage{xr-hyper}
\usepackage{multirow}
\usepackage{comment}
\usepackage{bm}
\usepackage[hidelinks=True]{hyperref}
\hypersetup{
  colorlinks   = true, 
  urlcolor     = blue, 
  linkcolor    = blue, 
  citecolor   = blue 
}
\begin{document}
 
\title{Quantized orbital and spin  Hall transport: interplay between  $sp$-hybridization, altermagnetism and spin-orbit coupling}

\author{Saikat Saha}
\affiliation{Department of Physics, BITS Pilani-Hyderabad Campus, Telangana 500078, India}
\affiliation{Department of Physics and Research Center OPTIMAS,
Rheinland-Pfälzische Technische Universität Kaiserslautern-Landau, 67663 Kaiserslautern, Germany}
\affiliation{Department of Physics, Indian Institute of Technology Bombay, Mumbai 400076, India}
\author{Banasree Sadhukhan}
\email{banasree.sadhukhan@mahindrauniversity.edu.in}
\affiliation{Department of Physics, Ecole Centrale School of Engineering, 
Mahindra University, Hyderabad, Telangana 500043, India}
\author{Tanay Nag}
\email{tanay.nag@hyderabad.bits-pilani.ac.in}
\affiliation{Department of Physics, BITS Pilani-Hyderabad Campus, Telangana 500078, India}

\begin{abstract}
We here explore the emergence of orbital and spin Hall effects, originating beyond the $L$-$S$ coupling, and investigate the interplay between inter-orbit hybridization,  relativistic Rashba spin-orbit coupling (SOC), and non-relativistic SOC, namely altermagnetic (AM) order, in a two-dimensional model Hamiltonian. The orbital (spin) Hall responses are remarkably found to be quantized within a window of Fermi energy when the strength of AM order (Rashba SOC) exceeds (falls below) the scale set by $sp$-hybridization. Importantly, orbital and spin Hall quantizations are independent of Rashba SOC and AM order, respectively, while the uniform profiles of finite orbital and vanishingly small spin moments of bands around the Fermi energy. The microscopic origin of such quantization comes from the Fermi surface-activated orbital and spin Berry curvatures. The extent of the quantized regime is strongly controlled by the intra-orbital coupling strength. As the temperature increases, the quantization is significantly compromised in the spin Hall case. We extend our analysis to the orbital and spin Nernst coefficients where the pronounced dip–peak structures signal the existence of the quantization leading to experimental relevance. 

\end{abstract}

\maketitle



\textit{\color{blue} Introduction-} The emergence of quantum  
Hall effect opens up a new era of condensed matter physics in terms 
of the electronic response due to an external magnetic field \cite{PhysRevLett.45.494}. A  voltage transverse to the probe electric field 
is developed while the Landau levels form due to the application of a magnetic field.  This has instigated the study of 
quantum metrology \cite{PhysRevLett.48.1559}, low-dissipation electronics \cite{weber20242024,collins2018electric}, quantum computation \cite{RevModPhys.80.1083},
and the topological quantum models and materials \cite{PhysRevLett.49.405,RevModPhys.82.3045,RevModPhys.83.1057,PhysRevLett.50.1395}. The quantum Hall effect persists in the absence of the external magnetic field, where the time-reversal symmetry is intrinsically broken due to inherent magnetization 
and leads to anomalous Hall (AH) effect \cite{RevModPhys.82.1959,RevModPhys.82.1539}. Over the years, a wide range of Hall phenomena has been identified, emerging from diverse combinations of external fields, intrinsic material properties, and the internal degrees of freedom of electrons. Beyond the conventional charge Hall effect in linear and non-linear regimes \cite{PhysRevB.107.245141,PhysRevB.104.115420,PhysRevB.106.045424,PhysRevB.105.214307,PhysRevB.107.L081110,PhysRevB.104.245122}, modern condensed matter physics now encompasses a broader class of Hall-like transport phenomena involving spin \cite{murakami2003dissipationless,PhysRevLett.92.126603,RevModPhys.87.1213}, orbital \cite{PhysRevLett.95.066601,PhysRevLett.121.086602}, valley \cite{PhysRevLett.99.236809,PhysRevB.77.235406},  thermal \cite{PhysRevLett.107.236601,PhysRevLett.104.066403,PhysRevLett.95.155901}, and skyrmionic \cite{PhysRevLett.107.136804} degrees of freedom.


Interestingly, the orbital Hall (OH) effect 
can arise in the absence of spin-orbit coupling (SOC) due to the  orbital texture where the momentum-dependent hybridization between different orbitals plays a very important role for producing a transverse flow of orbital angular momentum under an electric field \cite{gigantic,PhysRevLett.123.236403,PhysRevResearch.2.013177,PhysRevB.103.085113,PhysRevLett.83.1834,RevModPhys.87.1213,PhysRevLett.94.066602,PhysRevLett.102.016601,PhysRevB.98.214405,go2021orbitronics,burgos2024orbital,PhysRevLett.134.036304,PhysRevResearch.5.043052,PhysRevB.102.161103}.
On the other hand, in the presence of $L$-$S$ coupling of relativistic origin, causing the spin and orbital angular momentum to couple, the OH effect could transform into the spin Hall (SH) effect where up and down spins are separated in the transverse direction under the electric field that has been experimentally verified \cite{choi2023observation,wang2023inverse,santos2026probing}. The OH (SH) effect is a geometric transport phenomenon mediated by orbital (spin) Berry curvature that accounts for the winding of  internal orbital (spin) structure of Bloch states with the momentum \cite{PhysRevLett.95.066601}. Given the important role of SOC, the OH effect is found in two-dimensional transition metal dichalcogenides \cite{PhysRevB.101.121112,PhysRevB.102.035409,PhysRevB.101.161409,PhysRevLett.126.056601}. The OH effects are also proposed to explain large SH and AH effects in transition metal compounds \cite{PhysRevB.77.165117,PhysRevLett.100.096601,PhysRevLett.102.016601}. With the application of a thermal gradient instead of an electric field, orbital and spin Berry curvatures also cause orbital Nernst (ON) and  spin Nernst (SN) effect, respectively \cite{one,PhysRevB.106.024410}.

Very recently, an unconventional magnetic ordering, namely altermagnetism has been discovered experimentally \cite{PhysRevX.12.031042,PhysRevLett.133.206401,Reimers2024,PhysRevX.12.040002,PhysRevX.14.011019,PhysRevX.12.040501}, leading to the AH effect in the field-free situation \cite{PhysRevLett.133.086503,doi:10.1126/sciadv.aaz8809}.  Altermagnetism, originating from non-relativistic SOC,  breaks time-reversal symmetry without any net magnetization while the Fermi-surface simultaneously hosts spin-split
as well as spin-degenerate features of ferromagnetism and antiferromagnetism, respectively. The altermagnetic (AM) order promotes various non-trivial transport phenomena ranging from  diode effects \cite{ruthvik2025field,prnx-47mk,PhysRevB.111.L121401,pal2025topological}, to applications in spintronics \cite{RevModPhys.90.015005}. However, the consequences of AM order on OH and SH effects remain unexplored that we investigate in this study.


Given the above background on orbitronics and spintronics, the quantization of the Hall response is one of the central themes rather than the generation of the above transport coefficients. In this context, it would be extremely useful to study the interplay between momentum-dependent  inter-orbital and various spin-orbital couplings. In other words, can $sp$-hybridization, along with relativistic and non-relativistic SOC, stabilize the quantization phenomena for OH and SH effects within a window of Fermi energy? Considering two-dimensional model Hamiltonian without $L$-$S$ coupling, our study reveals that AM order (Rashba SOC) can induce a quantized  OH (SH) response when the 
magnitude of the above is higher (lower) than the strength of $sp$-hybridization. The window of the quantization strongly depends on the intra-orbit coupling strength while temperature significantly affects the SH quantization. The sharp dip-peak profiles of ON and SN coefficients are the direct manifestation of the quantized response which can be verified in experiments.

\textit{\color{blue}Model Hamiltonian--}
We consider two degrees of freedom, namely, spin and orbital angular momentum to compose the complete model Hamiltonian $\mathcal {H}(k_x,k_y)= \mathcal{H}_O (k_x,k_y)\otimes \mathcal{H}_S(k_x,k_y)$ in two dimensions. The orbital Hamiltonian $\mathcal{H}_O(k_x,k_y)$ is composed of the basis states  \(|s\rangle, |p_x\rangle, |p_y\rangle,  |p_z\rangle\) while the spin Hamiltonian $\mathcal{H}_S(k_x,k_y)$ is defined in the basis  of \(|\uparrow\rangle, |\downarrow\rangle\).  
We remove $k_x,k_y$ from the Hamiltonian for brevity. The non-zero matrix elements of  orbital Hamiltonian $\mathcal{H}_O$ are given by 
  $\langle s | \mathcal{H}_O | s \rangle = -\langle p_x | \mathcal{H}_O | p_x \rangle = -\langle p_y | \mathcal{H}_O | p_y \rangle= \langle p_z | \mathcal{H}_O | p_z \rangle=\Delta$, $ \langle s | \mathcal{H}_O | p_x \rangle
= -2i \gamma  \sin(k_y a) = \langle p_x  | \mathcal{H}_O | s\rangle^*$, and  $\langle s | \mathcal{H}_O | p_y \rangle
= -2i \gamma \sin(k_x a) = \langle p_y  | \mathcal{H}_O | s \rangle^*$  where the $\gamma$ is the strength of $sp_x$- and $sp_y$-hybridizations,   and $\Delta$ denotes the orbital energies \cite{PhysRevLett.121.086602}. On the  other hand,  the spin Hamiltonian is given by  $\mathcal{H}_S=t\big[\cos(k_x a) + \cos(k_y a)\big]\sigma_0 + J_A \big[\cos(k_x a) - \cos(k_y a)\big]\sigma_z + \lambda_R \sin(k_x a) \sigma_y + \lambda_R \sin(k_y a) \sigma_x $ where $t$ being the spin-independent hopping strength, \(J_A\) indicates  the  strength of AM order, and \(\lambda_R\) denotes the measure of Rashba SOC. $\sigma_{x,y}$ can mix the spin degrees of freedom similar to orbital mixing by $sp$-hybridization.
Note that the lattice spacing $a$ is considered to be unity henceforth.  We consider \(\Delta = 1 \) eV and \(\gamma = 0.25 \) eV  unless otherwise specified.  Note that we do not consider any direct $L$-$S$ coupling in the model.

\begin{figure}[t]
    \centering
\includegraphics[width=0.5\textwidth,angle=0]{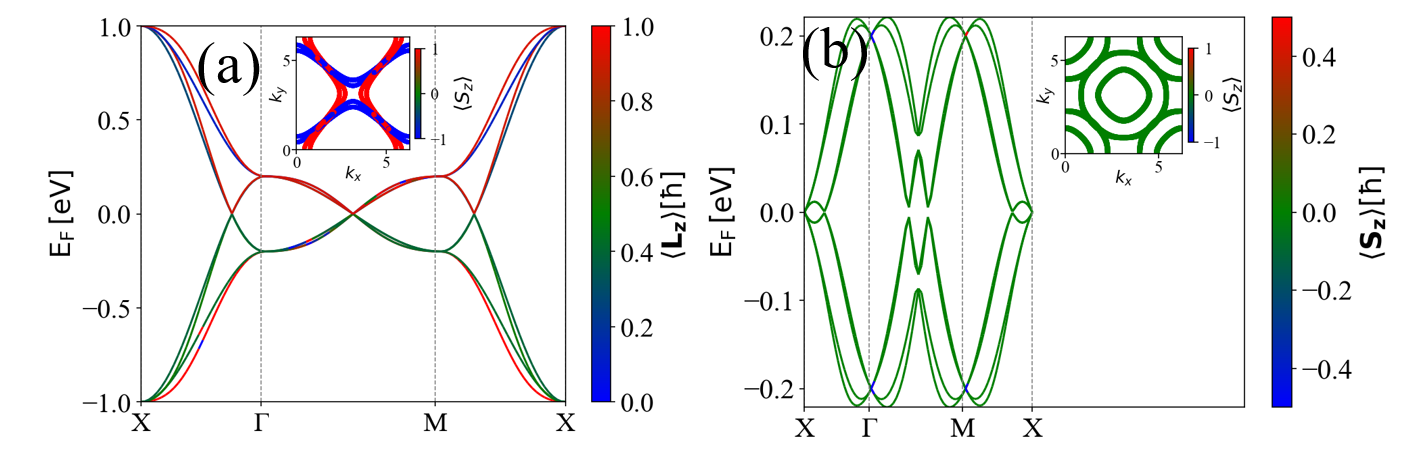}
    \caption{Band Structure (a) with the distribution of \( \langle L_z \rangle\) and (b) with the distribution of \( \langle S_z \rangle\), computed from the model Hamiltonian $\mathcal {H}$. The inset plot in (a) shows the Fermi surface contour for Fermi energy $E_F=0.1\, eV$ with distribution of \(\langle S_z \rangle \) in units of $\hbar$ and the inset plot of (b) shows the Fermi surface contour for Fermi energy $E_F=0.075\, eV$with distribution of \( \langle S_z \rangle\). We consider  \(\Delta = 1\, eV \) and \(\gamma = 0.25\, eV \) with $J_A = 1 \, eV, \lambda_R = 0\, eV$ in (a) and $J_A = 0 \, eV, \lambda_R = 0.1\, eV$ in (b).  
    }
    \label{fig:bands_all}
\end{figure}

We show the band dispersion along the high-symmetry lines in Figs. \ref{fig:bands_all} (a,b) for parameter set $(J_A,\lambda_R)=(1,0)\, eV$, and $(0,0.1)\, eV$, respectively. We find four-fold-degeneracy of bands at $\Gamma$ and $M$ points while the orbital weight, quantified by $\langle L_Z\rangle$ over each band, is higher towards the conduction bands as compared to the valence bands, Fig. \ref{fig:bands_all} (a). Interestingly, there exists eight-fold degeneracy at zero-energy between $\Gamma$ and $M$ points in the absence of SOC. We find spin-split bands at Fermi energy $E_F=0.1 \, eV$ as shown in the inset of Fig. \ref{fig:bands_all} (a) where AM nature is clearly visible with  $d$-wave profile. We further show that the completely spin-mixed energy bands in the presence of Rashba SOC while AM order is absent, see   Fig. \ref{fig:bands_all} (b). The concentric circle in the Fermi surface at $E_F=0.075 \, eV$  are caused by the spin-mixed quadratic bands in the presence of Rashba SOC, see inset of Fig. \ref{fig:bands_all} (b). Importantly, the $d$-wave profile is absent as $ J_A=0$. On the other hand, the degenerate points are shifted for $\lambda_R\ne 0$ along the high-symmetry axis, as expected, compared to the spin-polarized case with $\lambda_R=0$. However, the degree of degeneracies can be four-fold as well eight-fold at different momentum points which is universally observed for spin-polarized as well as spin-mixed cases.

\textit{\color{blue}Orbital and Spin Hall Effect--}
We now  proceed with the transport coefficients where orbital and spin Berry curvatures determines the OH and SH conductivities, respectively. 
The orbital and spin Berry curvature, associated with $n$-th band can be expressed as follows \begin{align}\label{eq:1}
\Omega_{n,\textbf{k}}^{z,\text{O(S)}}& = {\hbar^2}\sum_{m \neq n} \text{Im}\bigg[\frac{  \langle n_\textbf{k} | j_{y,\textbf{k}}^{z,\text{O(S)}} | m_\textbf{k} \rangle \langle m_\textbf{k} | v_{x,\textbf{k}}| n_\textbf{k} \rangle}
{(\epsilon_{m,\textbf{k}} - \epsilon_{n,\textbf{k}} +i\eta)^2} \bigg] \nonumber \\ &=\sum_{m \neq n} \Omega_{mn,\textbf{k}}^{z,\text{O(S)}}
\end{align}
where, \(|n_\textbf{k}\rangle, \epsilon_{n,\textbf{k}}\) are the $n$-th eigenvectors and eigenvalues of the Hamiltonian $\mathcal {H}$, respectively.  The internal orbital (spin) structure of Bloch states and its twisting in momentum space, is captured by the  orbital (spin) Berry curvature. The velocity operator is given by  $ v_{\alpha,\textbf{k}}= \frac{1}{\hbar}\frac{\partial\mathcal{H}}{ \partial k_\alpha}$ and $j_{\alpha,\textbf{k}}^{z,\text{O(S)}} = \frac{1}{2}\left\{ v_{\alpha,\textbf{k}}, \widetilde{L}_z(\widetilde{S}_z) \right\}$ where $\widetilde{L}_z = L_z\otimes\mathcal{I}_2$ and $\widetilde{S}_z=\mathcal{I}_4 \otimes \sigma_z$, see supplemental material (SM) for more details \cite{supp}. The band Berry curvature can be obtained when $j_{\alpha,\textbf{k}}^{z}$ is replaced with  $v_{\alpha,\textbf{k}}$. This orbital (spin) Berry curvature amplifies the contributions of different orbitals (spins)  around the degenerate  points yielding the features of a magnetic monopole similar to the band Berry curvature.  
At finite temperature with $T\ne 0$, the OH (SH) conductivity are given by  \cite{PhysRevLett.94.066602,PhysRevLett.102.016601}
\begin{align}\label{eq:3}
\sigma^{z}_{\rm OH(SH)}(T)
&=\frac{e}{\hbar} \sum_{m,n}^{m\ne n} \int\frac{d^2k}{(2\pi)^2}(f_{n,\textbf{k}}-f_{m,\textbf{k}})
\Omega_{mn,\textbf{k}}^{z,\text{O(S)}}
\end{align}
where, $f_{l,\textbf{k}} = \big(e^{\beta(\epsilon_{l,\textbf{k}}-E_F)}+1\big)^{-1}$, $\beta=(k_B T)^{-1}$, being the usual Fermi-Dirac distribution. Going beyond the response under the electric field, we also study the response under the temperature gradient. To be precise,  we examine the ON and SN conductivities as follows\cite{PhysRevMaterials.6.095001,PhysRevB.106.024410,PhysRev.181.1336} \\
\begin{equation}\label{eq:4}
   \sigma^{z}_{\rm ON(SN)} = \frac{\pi^2 k^2_B T}{-3e}\bigg[\frac{d}{dE} \sigma^{z}_{\rm OH(SH)}(T)\bigg]\Bigg|_{E=E_F}.
\end{equation}
We consider $N=500\times 500$ momentum grids for all our numerical calculations.

\begin{figure}[t]
    \centering
\includegraphics[width=0.45\textwidth,angle=0]{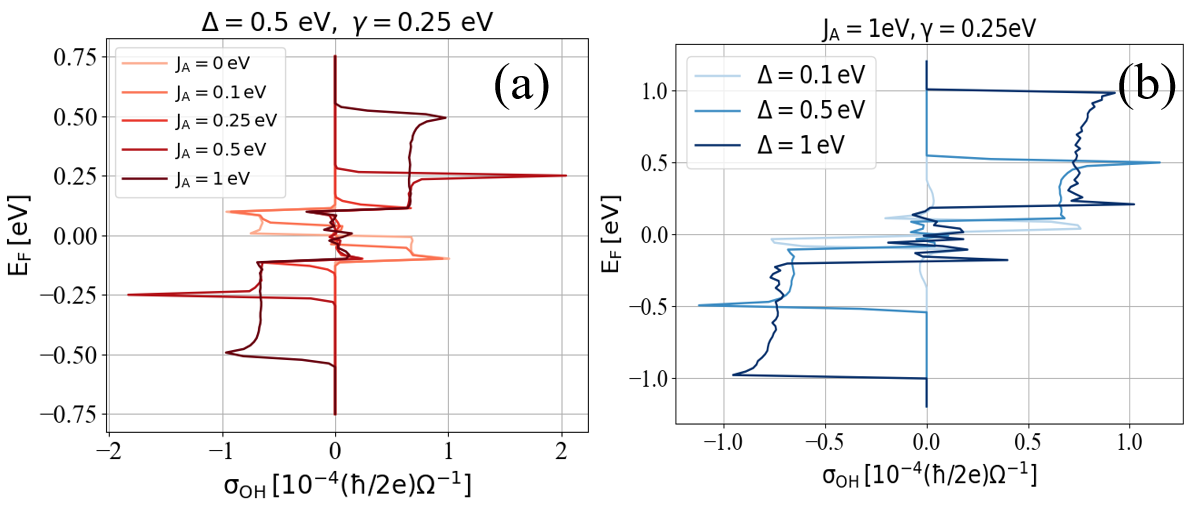}
    \caption{Variation of orbital Hall (OH) conductivities with Fermi energy $E_F$ is shown in (a) and (b), respectively, for different values of altermagnetic strength \(J_A\) and the onsite energy \(\Delta\). We consider $\lambda_R=0$. }
    \label{fig:ohct0}
\end{figure}

\textit{\color{blue}OH and SH conductivities at zero temperature--} We here study the OH response, following Eq. (\ref{eq:3}), caused by $sp$-hybridization in the absence of Rashba SOC $\lambda_R=0$.  The OH conductivity switches sign under the sign reversal of $E_F$ irrespective of the tuning parameters, exhibiting a universally odd behavior.
The OH conductivity is found to show a plateau-like structure with Fermi energy while the window of the above plateau increases with AM order $J_A$ above a critical strength $J_A^c \approx \gamma$, see Fig. \ref{fig:ohct0}(a). 
Interestingly, in the absence of AM order $J_A=0$, the OH response shows a tiny plateau that vanishes as AM order approaches the critical strength from $J_A<J_A^c$. Above this critical strength $J_A>J_A^c$, OH response switches its sign with Fermi energy. We find that these plateaus shrink and transform into a pronounced peak for $J_A \simeq J_A^c$; these peaks continue appearing at the boundaries of the plateau before the response vanishes for larger values of $E_F$. The outer boundary  of the plateau, obtained for $J_A<J_A^c$,  matches the inner boundary for $J_A>J_A^c$. 
The magnitude of OH response over these plateaus, irrespective of their extent, is qualitatively independent of $J_A$. This is clearly connected to an underlying quantized structure of the OH response.

This extended plateau profile, and its sign reversal with $E_F$ for $J_A > J_A^c$ can be microscopically explained by the Fermi surface contribution of the orbital Berry curvature. To be precise, the sign change of the response with the Fermi energy 
is a direct consequence of the  momentum 
distribution of the Fermi surface-activated orbital Berry curvature which is quantified by the integrand of Eq. (\ref{eq:3}), see SM for more details \cite{supp}. As is already known, any transport property is primarily mediated by the Fermi surface. The orbital Berry curvature around the degenerate points close to the Fermi surface contributes predominantly to the  OH conductivity. 
In the present case, the  quantity  $\big(f_{n,\mathbf{k}}-f_{m,\mathbf{k}}\big)\Omega_{mn,\mathbf{k}}^{z,\text{O}}$ effectively represents the 
distribution of orbital Berry curvature on the Fermi surface. The extent of the plateau is governed by the uniformity of the orbital Berry curvature on the Fermi surface over the momentum space. This distribution changes minimally with the Fermi energy, leading to a quantized outcome of the integral over momentum in Eq. (\ref{eq:3}). 
Interestingly,  this constant number over a certain window of $E_F$,  
essentially capturing a topological feature associated with the orbital degrees of freedom of the model.

It is also noteworthy that    $\langle L_z\rangle$ remains unaltered for a given band within a wide window of energy around the band edge. The shorter plateau for $J_A < J_A^c$ is also connected with a constant $\langle L_z\rangle$ behavior but at a narrower window of energy around the band center close to zero-energy. On the other hand, the peak is accompanied by a strong non-uniform profile of $\langle L_z\rangle$ at certain energies close to the band center. Therefore, the AM order can indeed control the orbital angular momentum profiles of the bands which results in a quantization of OH conductivity. 

We further check the robustness of the quantization by varying the on-site orbital energy $\Delta$. Upon increasing $\Delta$, the plateaus become extended in Fermi energy resulting in a larger quantization window, see Fig. \ref{fig:ohct0}(b). We consider $J_A>\gamma$ to highlight the quantization phenomenon. As explained earlier, the Fermi surface-activated orbital Berry curvature is expected to remain qualitatively unaltered when  $\Delta$ increases.
The on-site orbital energy tries to lift the degeneracy in the bands with a uniform distribution of $\langle L_z\rangle$ across a wide range of energy. This can be manifested through the plateau-like behavior while the fluctuations are more visible for larger $\Delta$ values. It is noteworthy that a slight increase in the quantized value is due to the increase in $\Delta$. The $sp$-hybridization and on-site energies play significant roles in determining the quantization window as well as its value.

The quantized Hall conductivity in the quantum case is directly related to the Landau levels that emerge under the application of a magnetic field \cite{PhysRevLett.45.494,PhysRevLett.49.405}. In the present case without any external magnetic field, Landau levels do not form, however, the bands exhibit uniform distribution of $\langle L_z\rangle$ over an energy window. The quantization continues to persist  
until the given energy level is occupied.   The quantization profile weakly depends on inter-orbit hybridization $\gamma$ and hopping $t$, see SM for more details \cite{supp}.    
Importantly, the Rashba SOC does not play any role in the OH effect.

\begin{figure}[t]
    \centering
\includegraphics[width=0.5\textwidth,angle=0]{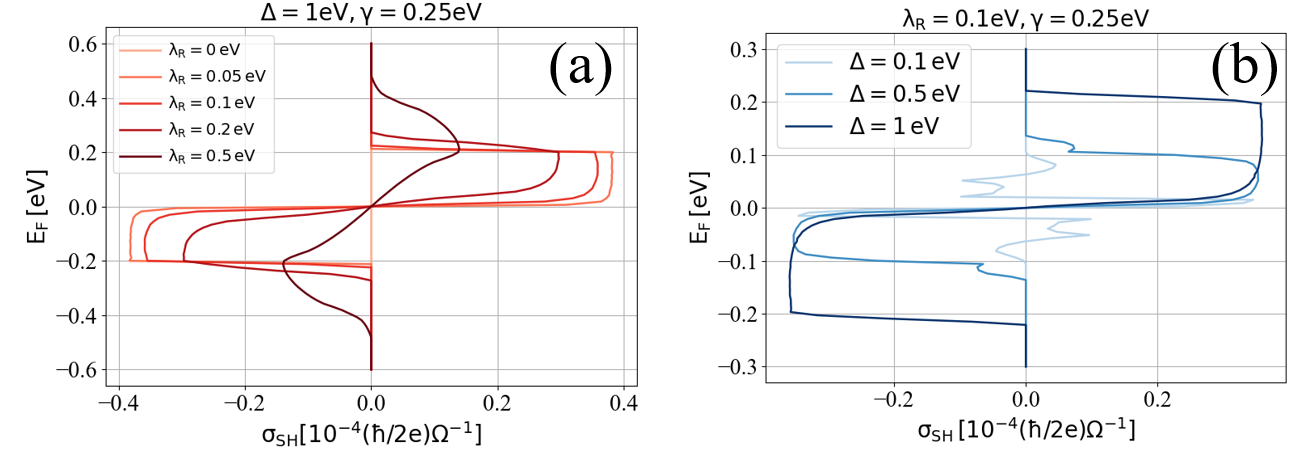}
    \caption{Variation of spin Hall (SH) conductivities with Fermi energy $E_F$ is shown in (a) and (b), respectively, for different values of SOC \(\lambda_R\) and the on-site orbital energy \(\Delta\). We consider $J_A=0$.
    }
    \label{fig:shct0}
\end{figure}

Having investigated the OH response, we now focus on SH response in the presence of Rashba SOC and $sp$-hybridization where the plateau structure clearly denotes the quantization. Unlike the previous case, the AM order is not mandatory to establish the quantization of  SH conductivity. 
We find a quantized response of SH conductivity for \(\lambda_R \ll \gamma\) while it vanishes in the absence of \(\lambda_R\), see Fig. \ref{fig:shct0}(a). The quantized value diminishes as \(\lambda_R \to \gamma\) and becomes completely non-quantized for  \(\lambda_R > \gamma\). Therefore, the quantization of SH conductivities strongly depends on $\lambda_R$  in terms of its quantitative values and qualitative profiles. Importantly, the window of the  plateau shortens with increasing $\lambda_R$.
Similar to the OH response,  the quantization and sign reversal can be explained by the spin Berry curvature profile over the Fermi surface, see SM for more details \cite{supp}. The robustness of the quantization window is governed by the largely unchanged momentum-space profile of the Fermi surface-activated spin Berry curvature. It is noteworthy that a uniform distribution of $\langle S_Z \rangle$ with vanishingly small value is also observed over a window of energy for different bands. Therefore, the quantization of SH conductivity might not be directly related to   $\langle S_Z \rangle$, unlike the correlation observed between a finite  value of $\langle L_Z \rangle$ and OH conductivity.

We extend our analysis in the context of $\Delta$ where the quantized value remains quantitatively unaltered while its window increases with $\Delta >\gamma$, see  Fig. \ref{fig:shct0}(b). As expected, increasing  $\Delta$ does not change the $\langle S_Z \rangle$ profiles of the bands, but rather stretches the energies of the bands, resulting in an almost identical profile of Fermi surface-activated Berry curvature. This causes an 
enlarged domain of Fermi energy, retaining the SH quantization. Interestingly, unlike the OH conductivity, the quantized value of the  SH response does not depend on $\Delta$. The quantization remains unaltered against the variation of inter-orbital hybridization $\gamma$ and hopping $t$, see SM for more details \cite{supp}.



\begin{figure}[t]
    \centering
\includegraphics[width=0.5\textwidth,angle=0]{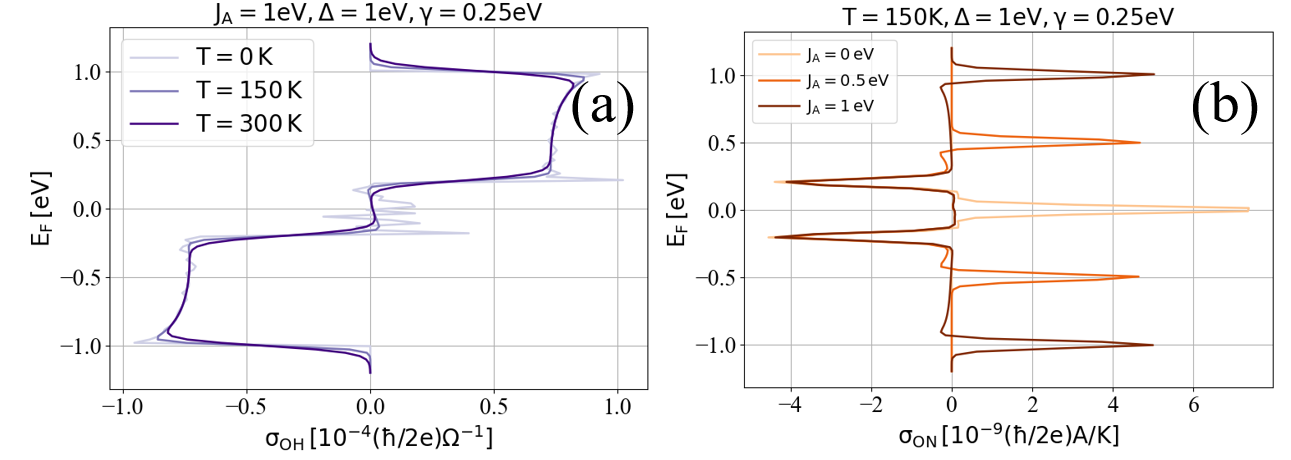}
    \caption{Variation of OH conductivity with $E_F$ for different values of temperature is shown in (a). The  Orbital Nernst (ON) conductivity for different values of \(J_A\) with respect to the Fermi energy is displayed in (b). 
    }
    \label{fig:ohct1}
\end{figure}

\textit{\color{blue}Conductivities at finite temperature--} We examine the effect of temperature on the quantization of OH conductivity, following Eq. (\ref{eq:3}), in Fig. \ref{fig:ohct1} (a). The zero-temperature fluctuations in OH coefficient around the boundaries of the plateau are flattened as the temperature rises without hampering the quantized profile. This leads to an experimentally measurable response at finite temperature. On the other hand, we examine ON conductivity under a temperature gradient, computed using Eq. (\ref{eq:4}),  in  Fig. \ref{fig:ohct1} (b) where the jumps of OH conductivity at the boundaries of the plateau are nicely captured by the strong peaks and dips at these values of $E_F$ and stays vanishingly small within the $E_F$ window enclosed by the boundaries. Interestingly, such a pronounced signal of ON conductivity can be considered as an experimental measure of the quantization window. The strong positive signal $E_F=0$ is a hallmark signature of $J_A=0$ and onset (destruction) of quantization yields in negative (positive) response for ON conductivity. The amplitude of the ON response does not change noticeably with $J_A$ resulting in identical amplitudes of ON conductivity. 
 In the context of quantum Hall effect, the above  behavior  is reminiscent of the longitudinal conductivity which shows peaks at the transition between the Hall plateaus.  Remarkably, the Nernst conductivity exhibits even-type of response as opposed to the odd-type of response for OH and SH conductivities.
Therefore, the complex effect of $sp$-hybridization and AM order is  manifested through the ON conductivity in the presence of a thermal gradient.

\begin{figure}[t]
    \centering
\includegraphics[width=0.5\textwidth,angle=0]{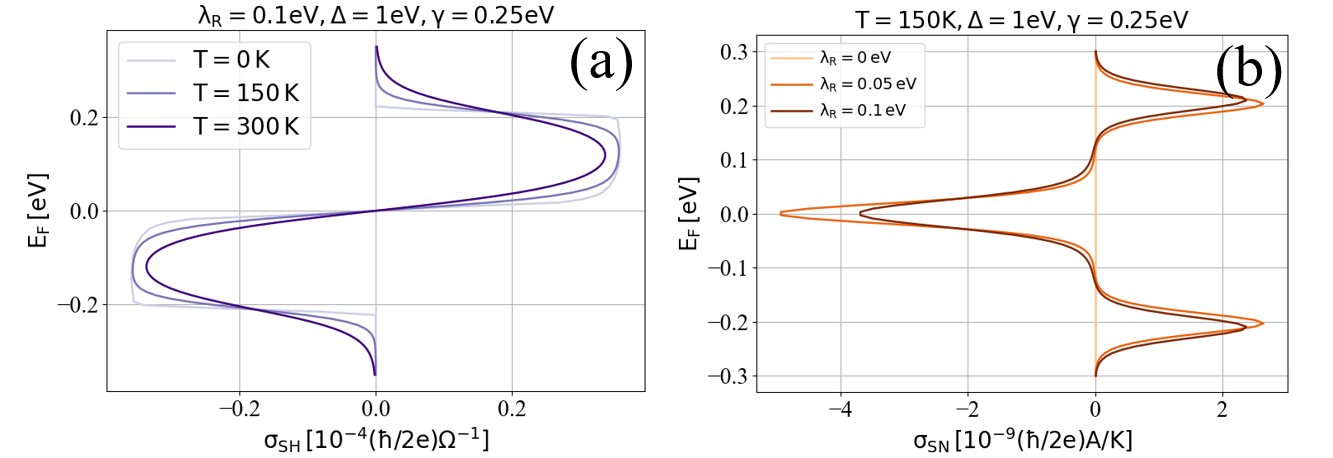}
    \caption{Variation of SH conductivity with $E_F$ for different values of temperature is shown in (a). The spin Nernst (SN) conductivity for different values of \(\lambda_R\) with respect to the Fermi energy is displayed in (b). 
    }
    \label{fig:shct1}
\end{figure}

We now focus on the finite temperature transport associated with spin degrees of freedom in Fig. \ref{fig:shct1}. Unlike the previous case, the quantization of SH conductivity breaks down as temperature increases. Therefore, the SH response is less immune to temperature than the OH conductivity, see Fig. \ref{fig:shct1} (a). The sharp jumps flattened leading to Gaussian-like profile of SH conductivity. On the other hand, the SN conductivity exhibits peaks and dips at non-zero and zero values of $E_F$, respectively, see Fig. \ref{fig:shct1} (b). The peaks (dips) are associated with the outer (inner) boundary of the plateau which is also similar to the ON conductivity, see Fig.  \ref{fig:ohct1} (b). The amplitude of the response diminishes as $\lambda_R$ increases which signifies the fact that the SH response is pronounced for a smaller value of Rashba SOC that is qualitatively similar to the quantization of  SH conductivity.

\textit{\color{blue}Discussions--} We now discuss the possible experimental relevance of our work where 
the SH response can be detected via nonlocal transport measurements and spin-torque techniques, similar to those used in heavy-metal systems \cite{Miron2011,doi:10.1126/science.1218197,PhysRevLett.98.156601,PhysRevLett.106.036601}. The OH effect, although not directly measurable, can be inferred through its conversion into spin accumulation via SOC, detectable using optical Kerr rotation \cite{PhysRevLett.119.087203}, inverse SH measurements \cite{PhysRevB.87.174417}, magnetic circular dichroism \cite{PhysRevB.50.12672, PhysRevX.12.031032,schattschneider2006detection}, leading to the possible experimental realization of our work.

We note  that the quantization of OH and SH coefficients does not depend on Rashba SOC and AM order, respectively, inferring the robustness of the responses against different parameters. Such a phenomenon 
can be utilized as a clear experimental knob where the interfacial engineering, determining the tuneability of Rashba SOC, and magnetic ordering control, experienced via strain, doping, or Neel vector manipulation, can be used to selectively access the OH and SH regimes. Therefore, the predicted quantized OH and SH responses can be probed in two-dimensional materials and thin-film heterostructures where orbital hybridization, Rashba SOC \cite{PhysRevB.105.195428,Yang2022,Wang2018}, and AM order \cite{Fu2025,PhysRevLett.133.056401} can be independently tuned. Candidate platforms include transition-metal thin films, oxide interfaces, and recently identified AM systems such as RuO$_2$ and MnTe \cite{Amin2024,doi:10.1126/sciadv.adj4883}.

\textit{\color{blue}Conclusions--} We investigate the emergence of OH and SH responses beyond conventional $L$-$S$ coupling within a two-dimensional model Hamiltonian, focusing on the interplay between inter-orbital hybridization, relativistic Rashba SOC, and non-relativistic spin splitting arising from AM order.
We demonstrate that the OH (SH)  conductivity becomes quantized over a finite Fermi energy window when the strength of AM order (Rashba SOC) exceeds (remains below) the characteristic inter-orbit $sp$-hybridization scale. Notably, the quantization of the OH (SH) response is insensitive to Rashba SOC (AM order), reflecting two distinct underlying mechanisms behind OH and SH responses.
The Fermi surface-activated orbital and spin Berry curvatures primarily contribute to the quantized response. 
The quantization  is 
further correlated with nearly uniform distributions of finite orbital and vanishingly small spin moments in the bands near the Fermi level. The width of the quantized regime is strongly governed by the intra-orbital coupling strength. Finite-temperature effects significantly degrade the quantization, particularly for the spin Hall response. Extending our analysis to thermoelectric transport, we find that the ON and SN coefficients exhibit pronounced dip–peak structures, providing clear experimental signatures of the underlying quantized Hall responses.


\textit{\color{blue}Acknowledgments--} TN and SS thank  BITS Pilani for the NFSG Grant No. NFSG/HYD/2023/H0911. BS  thanks the Prime Minister’s Early
Career Research Grant (PMECRG) of the Anusandhan National Research Foundation Grant No.
ANRF/ECRG/2024/005021/PMS. TN thanks the Advanced Research Grant (ARG) from Anusandhan National Research Foundation Grant No. ANRF/ARG/2025/003163/PS. 

\bibliographystyle{apsrev4-1}
\bibliography{reference}

\normalsize\clearpage

\begin{onecolumngrid}
	\begin{center}
    
		{\fontsize{12}{12}\selectfont
        
			\textbf{Supplemental Material for ``Quantized orbital and spin  Hall transport: interplay between  $sp$-hybridization, altermagnetism and spin-orbit coupling''\\[5mm]}}
            
		{\normalsize  Saikat Saha,$^{1,2,3}$, Banasree Sadhukhan,$^4$, and  Tanay Nag$^{1}$, \\
        {\small $^1$\textit{Department of Physics, BITS Pilani-Hyderabad Campus, Telangana 500078, India}}\\[0.5mm]}
		{\small $^2$\textit{Department of Physics and Research Center OPTIMAS,
        Rheinland-Pfälzische Technische Universität Kaiserslautern-Landau, 67663 Kaiserslautern, Germany}\\[0.5mm]}{\small $^3$\textit{Department of Physics, Indian Institute of Technology Bombay, Mumbai 400076, India}\\[0.5mm]}
        
		{\small $^4$\textit{Department of Physics, Ecole Centrale School of Engineering, 
Mahindra University, Hyderabad, Telangana 500043, India}}
		{}
	\end{center}
	
	\newcounter{defcounter}
	\setcounter{defcounter}{0}
	\setcounter{equation}{0}
	\renewcommand{\theequation}{S\arabic{equation}}
	\setcounter{figure}{0}
	\renewcommand{\thefigure}{S\arabic{figure}}
	\setcounter{page}{1}
	\pagenumbering{roman}
	
	\renewcommand{\thesection}{S\arabic{section}}
	
	

\section{Orbital representation of the Hamiltonian}
\label{SM1}

We now express the Hamiltonian in terms of the orbital angular momentum. We demonstrate below the angular momentum matrix in terms of the basis sets of $|s\rangle$, $|p_x\rangle$, $|p_y\rangle$, and $|p_z\rangle$. One can cast the quantum states $|l,m\rangle$ in terms of  azimuthal quantum number $l$ and the magnetic quantum number $m$. The orbital angular momentum operator is given by $\textbf {L}=(L_x,L_y,L_z)$ while the raising and lowering operators are given by $L_x = \frac{1}{2}\left(L_+ + L_-\right)$, $L_y = \frac{1}{2i}\left(L_+ - L_-\right)$. The operations are defined as follows $L_\pm|l,m\rangle = \hbar\sqrt{l(l+1)-m(m\pm1)}|l,m\pm1\rangle$ and   $L_z|l,m\rangle = m\hbar|l,m\rangle$. Considering the states $|l,m\rangle$ with $m=-l,-l+1,\cdots, l-1,l$ where $l=0,1$, and  the basis states are explicitly given by $|s\rangle=|0,0\rangle$, $|p_x\rangle = \frac{1}{\sqrt{2}} \left( |1,-1\rangle - |1,1\rangle \right)$, and $|p_y\rangle = \frac{i}{\sqrt{2}} \left( |1,-1\rangle + |1,1\rangle \right)$, and  $|p_z\rangle=|1,0\rangle$. Therefore, the $L_x,L_y,L_z$ matrices  in the basis $|s\rangle$, $|p_x\rangle$, $|p_y\rangle$, and $|p_z\rangle$ are given by \\
$
L_x = \hbar
\begin{pmatrix}
0 & 0 & 0 & 0\\
0 & 0 & 0 & 0\\
0 & 0 & 0 & -i\\
0 & 0 & i & 0\\
\end{pmatrix}
$,
$
L_y = \hbar
\begin{pmatrix}
0 & 0 & 0 & 0\\
0 & 0 & 0 & i\\
0 & 0 & 0 & 0\\
0 & -i & 0 & 0\\
\end{pmatrix}
$, and 
$
L_z = \hbar
\begin{pmatrix}
0 & 0 & 0 & 0\\
0 & 0 & -i & 0\\
0 & i & 0 & 0\\
0 & 0 & 0 & 0\\
\end{pmatrix}
$. 

Now after introducing the spin-degrees of freedom in the basis \(|\uparrow\rangle, |\downarrow\rangle\), one has to use the Pauli matrices $\sigma_x$, $\sigma_y$, and $\sigma_z$. Therefore, to invoke the orbital and spin degrees of freedom in the Hamiltonian, the basis states are given by $|s\uparrow\rangle$, $|p_x\uparrow\rangle$, $|p_y\uparrow\rangle$, $|p_z\uparrow\rangle$, $|s\downarrow\rangle$, $|p_x\downarrow\rangle$, $|p_y\downarrow\rangle$ and $|p_z\downarrow\rangle$.  As a result, the Hamiltonian can be expressed in terms of the operators $L_{i}\otimes \sigma_j$, with $i,j=x,y,z$. 
The orbital Hamiltonian $\mathcal{H}_l(k_x,k_y)$ is composed of the basis states  \(|s\rangle, |p_x\rangle, |p_y\rangle,  |p_z\rangle\) while the spin Hamiltonian $\mathcal{H}_s(k_x,k_y)$ is defined in the basis  of \(|\uparrow\rangle, |\downarrow\rangle\). Therefore, the total Hamiltonian is written in the basis $|s \uparrow \rangle, |p_x \uparrow\rangle, |p_y\uparrow\rangle,  |p_z \uparrow\rangle, |s\downarrow\rangle, |p_x \downarrow\rangle, |p_y \downarrow\rangle,  |p_z \downarrow\rangle$. We consider spherical harmonic basis $Y_l^m(\theta,\phi)$ with $l=1$ and $m=-1,0,1$ to represent the  $|p_{x,y,z}\rangle$ orbitals while $|s\rangle$ is associated with   spherical harmonic basis $Y_0^0(\theta,\phi)$.

In the similar line, $|d_{x^2-y^2,z^2,xy,yz,xz}\rangle$ are originated from  $Y_2^m(\theta,\phi)$ with $m=-2,-1,0,1,2$. To be precise, $|d_{x^2-y^2}\rangle = \frac{1}{\sqrt{2}} \left( |2,2\rangle + |2,-2\rangle \right)$, $|d_{xy}\rangle = -\frac{i}{\sqrt{2}} \left( |2,2\rangle - |2,-2\rangle \right)$, $|d_{xz}\rangle = \frac{1}{\sqrt{2}} \left( |2,1\rangle - |2,-1\rangle \right)$,
$|d_{yz}\rangle = -\frac{i}{\sqrt{2}} \left( |2,1\rangle + |2,-1\rangle \right)$,
and  $|d_{z^2}\rangle=|2,0\rangle$. One can obtain  the following form of $L_x$, $L_y$ and $L_z$ in the basis $|d_{xy}\rangle$, $|d_{yz}\rangle$, $|d_{xz}\rangle$, $|d_{x^2-y^2}\rangle$, and $|d_{z^2}\rangle$ are given by 
$\\
L_x = \hbar
\begin{pmatrix}
0 & 0 & 0 & 0 & \sqrt{3}\\
0 & 0 & i & 0 & 0\\
0 & -i & 0 & 0 & 0\\
0 & 0 & 0 & 0 & 1\\
\sqrt{3} & 0 & 0 & 1 & 0\\
\end{pmatrix}
$,
$
L_y = \hbar
\begin{pmatrix}
0 & 0 & -i & 0 & 0\\
0 & 0 & 0 & 0 & i\sqrt{3}\\
i & 0 & 0 & -i & 0\\
0 & 0 & i & 0 & 0\\
0 & -i\sqrt{3} & 0 & 0 & 0\\
\end{pmatrix}
$, and 
$
L_z = \hbar
\begin{pmatrix}
0 & 2i & 0 & 0 & 0\\
-2 i & 0 & 0 & 0 & 0\\
0 & 0 & 0 & -i & 0\\
0 & 0 & i & 0 & 0\\
0 & 0 & 0 & 0 & 0\\
\end{pmatrix}
$. 

\section{Microscopic origin of quantization}
\label{SM4}

\begin{figure*}[ht]
    \centering
\includegraphics[scale=0.5]{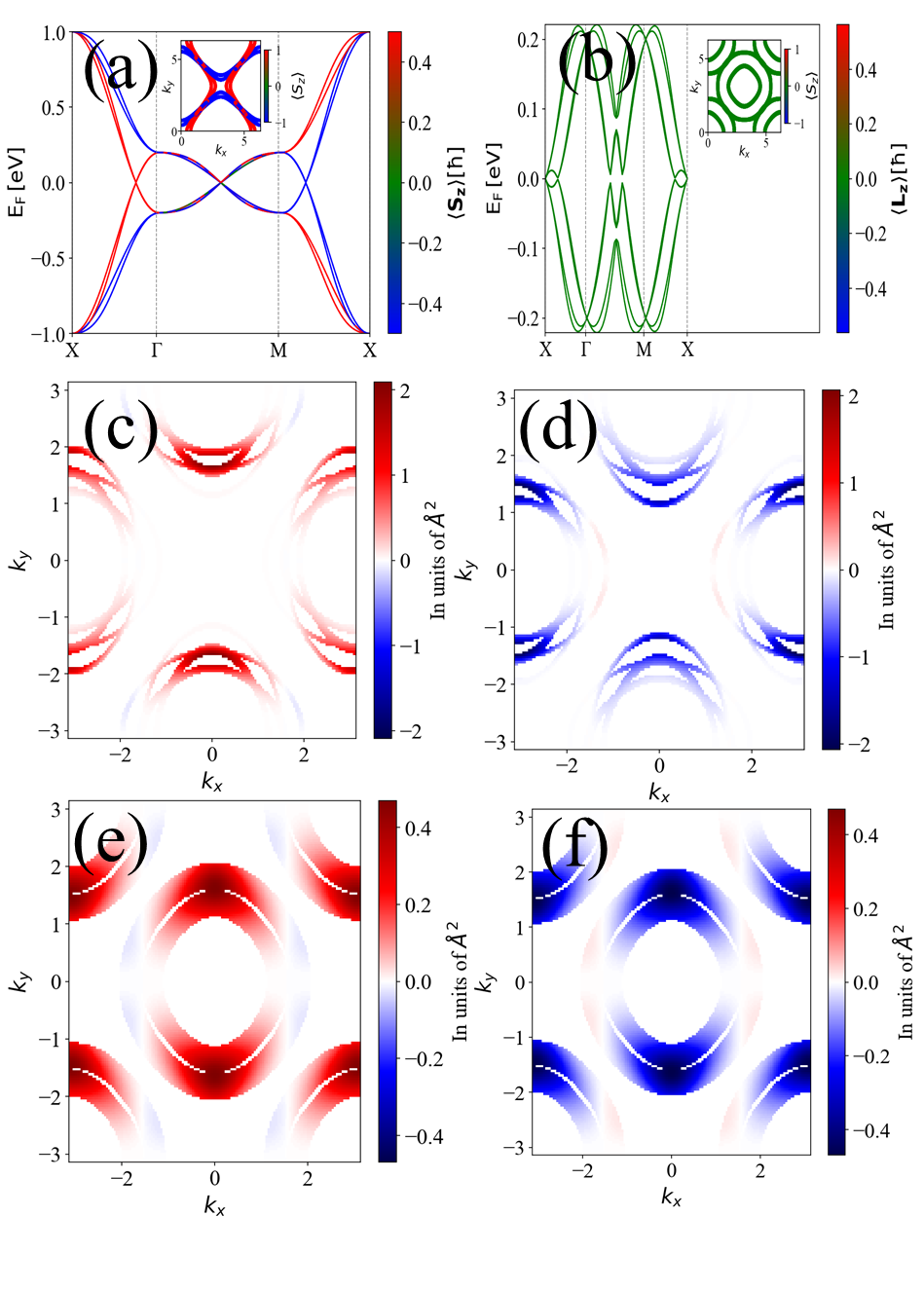}
    \caption{Band Structure (a) with the distribution of \( \langle S_z \rangle\) for $E_F=0.1 \, eV$ and (b) with the distribution of \( \langle L_z \rangle\) for $E_F=0.075\, eV$. 
    The other parameters are \(\Delta = 1\, eV \)  and \(\gamma = 0.25\, eV \) with $J_A = 1 \, eV, \lambda_R = 0\, eV$ in (a) and $J_A = 0 \, eV, \lambda_R = 0.1\, eV$ in (b) respectively. The integrand $\big(f_{n,\mathbf{k}}-f_{m,\mathbf{k}}\big)\Omega_{n,\mathbf{k}}^{z,\text{O[S]}}$ in Eq. (2) of the main text, capturing the Fermi surface contribution of   
    orbital [spin] Berry curvature,  are shown in (c,d) and [(e,f)] for $E_F=0.3\, eV$ and $-0.3\, eV$ [$E_F=0.1\, eV$ and $-0.1\, eV$], respectively. We consider  $\Delta = 0.5 \, eV,\gamma = 0.25 \, eV, J_A = 1 \, eV, \lambda_R = 0 \,eV$ for (c,d). We choose $\Delta = 1 \, eV,\gamma = 0.25 \, eV, J_A = 0 \, eV, \lambda_R = 0.05 \,eV$ for (e,f). 
    }
    \label{fig:bands_berry_curvatures}
\end{figure*}

We show the distribution of spin weight, captured by $\langle S_z \rangle$, of the individual bands in Fig. \ref{fig:bands_berry_curvatures}(a)
where the effect of AM order is clearly manifested through the spin-split as well as the spin-degenerate natures of the bands. To examine the effect of altermagnetism, we consider \(\Delta = 1\, eV \), \(\gamma = 0.25\, eV \) with $J_A = 1 \, eV, \lambda_R = 0\, eV$ as taken for Fig.  1(a) of the main text.  
The inset of Fig. \ref{fig:bands_berry_curvatures}(a) depicts the Fermi surface $C_4$ symmetry of the Fermi surface.  On the other hand, we show the $\langle L_z \rangle$ of the bands in Fig. \ref{fig:bands_berry_curvatures}(b) where vanishingly small  distribution of 
$\langle L_z \rangle$ signifies the absence of orbital ordering in the presence of Rashba SOC only. 
In the above, to emphasize the effect of Rashba SOC,  we consider \(\Delta = 1\, eV \), \(\gamma = 0.25\, eV \) with $J_A = 0 \, eV, \lambda_R = 0.1\, eV$, as taken for  Fig. 1(b) of the main text. The inset of  Fig. \ref{fig:bands_berry_curvatures}(b) shows the Fermi surface of concentric rings, which are reminiscent of the effect originating from Rashba SOC.

In order to understand the microscopic origin of quantization, one can study the integrand of Eq. (2) of the main text.  Note that any transport property is mainly governed by the Fermi surface. In other words, the Fermi surface selects the relevant momentum modes which 
contribute to the integral of Eq. (2) the most. The contribution of Berry curvature over the Fermi surface is the key ingredient and yields the maximum contribution to the integral.   To be precise, the quantity  $\big(f_{n,\mathbf{k}}-f_{m,\mathbf{k}}\big)\Omega_{mn,\mathbf{k}}^{z,\text{O(S)}}$ effectively reduces to the form $\sum_{m \neq n} \big( \frac{\delta f_{mn}}{\delta \epsilon_{mn}}\big)  \text{Im}\big[\frac{  \langle n_\textbf{k} | j_{y,\textbf{k}}^{z,\text{O(S)}} | m_\textbf{k} \rangle \langle m_\textbf{k} | v_{x,\textbf{k}}| n_\textbf{k} \rangle}
{(\epsilon_{m,\textbf{k}} - \epsilon_{n,\textbf{k}} +i \eta)} \big]$ where the first term $\big( \frac{\delta f_{mn}}{\delta \epsilon_{mn}}\big)$ refers to the Dirac-delta function picking up the Fermi momentum only while second term carries the signature of the orbital or spin Berry curvature. Here, $f_{m,\mathbf{k}}-f_{n,\mathbf{k}}=\delta f_{mn}$ and $\epsilon_{m,\textbf{k}} - \epsilon_{n,\textbf{k}}=\delta \epsilon_{mn}$Therefore, the integrand  can be written in the following form $\sum_{m \neq n} \big(f_{n,\mathbf{k}}-f_{m,\mathbf{k}}\big)\Omega_{mn,\mathbf{k}}^{z,\text{O(S)}} \approx \sum_{n} \delta(\textbf{k}-k_F) \text{Im}\big[\frac{  \langle n_\textbf{k} | j_{y,\textbf{k}}^{z,\text{O(S)}} | n_\textbf{k} \rangle \langle n_\textbf{k} | v_{x,\textbf{k}}| n_\textbf{k} \rangle}
{(E_F - \epsilon_{n,\textbf{k}} + i \eta )} \big]  $. This directly indicates that the denominator of the second term diverges at the Fermi energy, causing a significant contribution to the integral and the numerator has to be evaluated for 
$\textbf{k}$-modes lying over the Fermi surface only. It is important to note that the summation of $m$ is absorbed by the quantity $\big( \frac{\delta f_{mn}}{\delta \epsilon_{mn}}\big) $ that accounts for the Fermi surface. Therefore,  the quantity $\text{Im}\big[\langle n_\textbf{k} | j_{y,\textbf{k}}^{z,\text{O(S)}} | n_\textbf{k} \rangle \langle n_\textbf{k} | v_{x,\textbf{k}}| n_\textbf{k} \rangle \big]$ is very important to determine the sign of the response after the momentum integration in Eq. (2) of the main text. The Fermi surface-activated Berry curvature i.e., $\sum_{n} \delta(\textbf{k}-k_F) \text{Im}\big[\frac{  \langle n_\textbf{k} | j_{y,\textbf{k}}^{z,\text{O(S)}} | n_\textbf{k} \rangle \langle n_\textbf{k} | v_{x,\textbf{k}}| n_\textbf{k} \rangle}
{(E_F - \epsilon_{n,\textbf{k}} + i \eta )} \big] $, derived from the integrand of Eq. (2) of the main text, is the main contributor. 
Given the above heuristic analysis, we numerically explicitly compute   $\big(f_{n,\mathbf{k}}-f_{m,\mathbf{k}}\big)\Omega_{mn,\mathbf{k}}^{z,\text{O(S)}}$, capturing the Fermi surface contribution of the orbital and spin Berry curvatures in Figs. \ref{fig:bands_berry_curvatures} (c,d) and (e,f), respectively. To highlight the sign change of the response, as shown in Figs. 2(a) and 3(a) of the main text, we compute the above quantity with the Fermi energy of $E_F=0.3\, eV$ and $-0.3\, eV$ [$E_F=0.1\, eV$ and $-0.1\, eV$] for orbital [spin] Hall response. The momentum distributions are exactly opposite of each other between positive and negative values of $E_F$. 
The overall positive contribution for the positive Fermi energy and negative contribution for the negative Fermi energy with the same magnitudes clearly explain the sign flip of Hall responses in the positive and negative regions of Fermi energy, as shown in Figs. 2 and 3 of the main text.


\section{Robustness of OH quantization}
\label{SM2}

\begin{figure*}[t]
    \centering
\includegraphics[scale=0.5]{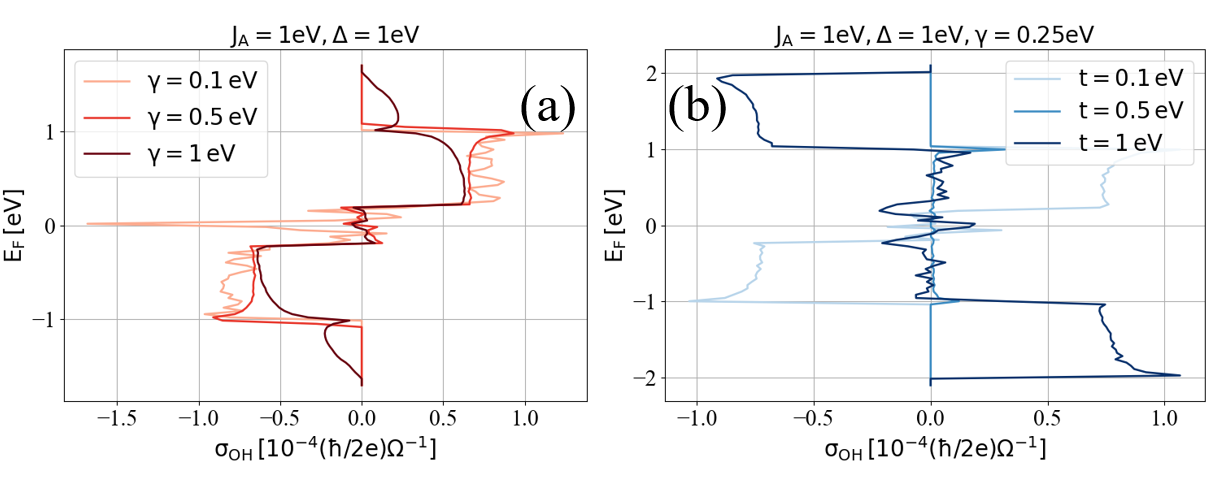}
    \caption{Variation of Orbital Hall Conductivity (\textbf{a}) for different values of \(\gamma\) and (\textbf{b}) for different values of t with respect to fermi energy
    }
    \label{fig:ohcgamma1}
\end{figure*}


We show the behavior of OH conductivity with Fermi energy $E_F$ for different values of $\gamma$, see  Fig. \ref{fig:ohcgamma1}(a). For smaller $sp$-hybridization $\gamma=0.1$eV, the response shows fluctuations around a mean value. After increasing $\gamma$, the fluctuations reduce, and OH conductivity shows quantization in terms of plateau-like region. Further increasing $\gamma$, the plateau structure starts dissolving and leading to a breakdown of  quantization. Therefore, the quantization profile strongly depends on the value of the $sp$-hybridization, which has to be less than the AM strength $J_A$. The AM order helps in stabilizing the quantization while the $sp$-hybridization determines $\langle L_z \rangle$ profiles over the bands. We examine the effect of the variation of hopping with the quantization, see Fig. \ref{fig:ohcgamma1}(b). The energy of the bands increases as hopping strength increases. This results in a wider window of quantization while the inner boundary of the plateau shifts to higher Fermi energy. However, quantization is destroyed for sufficiently small values of hopping strength. The quantization continues to persist once hopping leads to dispersive bands where degeneracy and non-degeneracy are both present.


\section{Robustness of SH quantization}
\label{SM2}

\begin{figure*}[t]
    \centering
\includegraphics[scale=0.5]{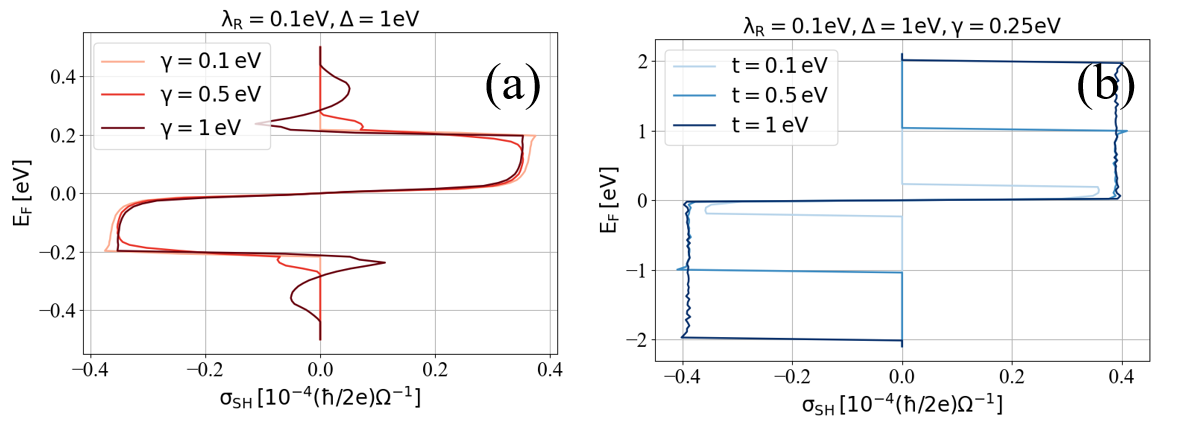}
    \caption{Variation of Spin Hall Conductivity (\textbf{a}) for different values of \(\gamma\) and (\textbf{b}) for different values of t with respect to fermi energy
    }
    \label{fig:shcgamma1}
\end{figure*}


We show the variation of SH conductivity with $E_F$ for different values of $\gamma$, see  Fig. \ref{fig:shcgamma1} (a). The plateau-like structure remains unaltered with $\gamma$ while a non-zero value is required to observe the quantization. The magnitude of $sp$-hybridization has to be larger than the Rashba SOC strength leading to the quantization.  The spin expectation $\langle S_z\rangle$  shows a uniform profile over bands and does not change with $\gamma$. We next focus on the variation of SH conductivity with $E_F$ 
for different values of $t$, see  Fig. \ref{fig:shcgamma1} (b). The energy of the bands increases as $t$ increases and yields a wider plateau profile. The SH response remains quantized at the same  value as hopping increases. The spin expectation $\langle S_z\rangle$ does not change as $t$ changes, inferring the fact that the SH response is more stable than the OH response.

\end{onecolumngrid}

\end{document}